\documentclass[11pt]{article}
\pdfoutput=1
\usepackage[utf8]{inputenc}
\usepackage[margin = 1in]{geometry}
\usepackage{amsfonts, amssymb, amsmath, amsthm}
\usepackage{hyperref}
\usepackage{cleveref}
\usepackage{comment}
\usepackage{mathtools}
\usepackage{pdfpages}
\allowdisplaybreaks
\usepackage{bm}
\usepackage{enumerate}
\usepackage{graphicx}
\usepackage{float}
\usepackage{xcolor}
\usepackage{array}
\usepackage{multicol, multirow}
\hypersetup{
    colorlinks = true,
    urlcolor  = blue,
    citecolor = blue,
}

\newcommand{\diff}[1]{\dfrac{d{#1}}{dt}}
\newcommand{\Prob}{\mathbb{P}}

\usepackage{tikz}
\usetikzlibrary{shapes.geometric, arrows, quotes}
\tikzstyle{S_node} = [circle, radius = 1cm, draw=black, fill=yellow!30]
\tikzstyle{Ip_node} = [circle, radius = 1cm, draw=black, fill=red!30]
\tikzstyle{It_node} = [circle, radius = 1cm, draw=black, fill=red!60]
\tikzstyle{R_node} = [circle, radius = 1cm, draw=black, fill=green!60]
\tikzstyle{E_node} = [circle, radius = 1cm, draw=black, fill=gray!30]
\tikzstyle{U_node} = [circle, radius = 1cm, draw=black, fill=green!30]
\tikzstyle{In_node}= [rectangle, minimum width=3cm, minimum height=1cm, draw=black, fill=blue!30]
\tikzstyle{Mn_node}= [rectangle, minimum width=3cm, minimum height=1cm, draw=black, fill=yellow!30]
\tikzstyle{Mp_node}= [rectangle, minimum width=3cm, minimum height=1cm, draw=black, fill=red!30]
\tikzstyle{Mi_node}= [rectangle, minimum width=3cm, minimum height=1cm, draw=black, fill=red!60]
\tikzstyle{Out_node}=[rectangle, minimum width=3cm, minimum height=1cm, draw=black, fill=yellow!60]

\tikzstyle{arrow} = [thick,->,>=stealth]

\usepackage[symbol]{footmisc}

\usepackage{authblk}
\title{A Generalized Epidemiological Model for COVID-19 with Dynamic and Asymptomatic Population}

\author[1]{Anirban Ghatak}
\author[2]{Shivshanker Singh Patel\footnote{Corresponding author: Decision Science, Indian Institute of Management Visakhapatnam, Andhra Pradesh, 530003, INDIA \\ Email: shivshanker@iimv.ac.in}}
\author[3]{Soham Bonnerjee}
\author[3]{Subhrajyoti Roy}
\affil[1]{IIM Kozhikode, INDIA}
\affil[2]{IIM Visakhapatnam, INDIA}
\affil[3]{ISI Kolkota, INDIA}

\begin{document}
\maketitle

\begin{abstract}
In this paper, we develop an extension of standard epidemiological models, suitable for COVID-19. This extension incorporates the transmission due to pre-symptomatic or asymptomatic carriers of the virus. Furthermore, this model also captures the spread of the disease due to the movement of people to/from different administrative boundaries within a country. The model describes the probabilistic rise in the number of confirmed cases due to the concomitant effects of (incipient) human transmission and multiple compartments.  The associated parameters in the model can help architect the public health policy and operational management of the pandemic. For instance, this model demonstrates that increasing the testing for symptomatic patients does not have any major effect on the progression of the pandemic, but testing rate of the asymptomatic population has an extremely crucial role to play. The model is executed using the data obtained for the state of Chhattisgarh in the Republic of India. The model is shown to have significantly better predictive capability than the other epidemiological models. This model can be readily applied to any administrative boundary (state or country). Moreover, this model can be applied for any other epidemic as well.
\end{abstract}

\section{Introduction}
\subsection{The Global Scene}
The Novel Coronavirus, or SARS-CoV-2, or as it is commonly called, COVID-19, is spreading throughout the globe in a rapid pace, and it has already caused destruction of an unprecedented scale, economically, physically, and socially. The reproducibility of the virus, the proportion of asymptomatic carrier, the absence of antibodies of the virus in human bodies, and most importantly, the lack of experience of the people in general in handling such a scenario has majorly contributed to this catastrophe. At the time of writing this paper, around 29 millions of people are infected throughout the world, and around 1 million people has already died of the virus. Although the scale of mortality may not sound extremely catastrophic considering the world population of 7.8 billion, but when one considers this to happen in a time period of 8 months, without showing any signs of slowing down in most of the world, it becomes a matter of extreme concern and emergency.

A prediction is always an extremely difficult problem in situations like these where enough data points are not available and there is no history of an outbreak of contagious disease in this scale. While, in general, standard methods of learning from earlier outbreaks or outbreaks in other countries can give the researchers a decent head-start, due to various policies adopted by various governments, tracking the spread of the virus in different countries as a learning mechanism does not prove to be a fruitful task. 

\subsection{Background: The case of India}

The first documented case of SARS-COV-2 in India was reported on 30 January 2020 in the state of Kerala. As per the current count, India has the largest number of confirmed cases in Asia, and has the second highest number of confirmed cases in the world after the United States \cite{hannah_2020}. The total number of confirmed cases in India crossed the 100,000 mark on 19 May, 200,000 on 3 June, and 1,000,000 confirmed cases on 17 July 2020 \cite{covid19indiaorg2020tracker}. The documented mortality rate in India is among the lowest in the world at around 1.8\% as of 29 August 2020 and is showing a monotonically decreasing trend \cite{covid19indiaorg2020tracker}.

India has gone through several restrictive phases in order to contain the spread of the virus. The first nationwide lockdown was announced on 24 March 2020 for 21 days, and that stopped all modes of transportation within and outside of the country. On 14 April, India extended the lockdown till 3 May and that was further continued by two-week extensions starting 3 and 17 May. From 3 May 2020 onwards, the government of India has started to relax the restrictions of lockdown and allowed migrant workers to return to their home state, officially, for the first time since 24 March 2020. This is an important moment for this paper, and we have deliberated on this event later. From 1 June, the government started ``unlocking" the country (barring ``containment zones") in three unlock phases.

The ability to predict the spread of virus beforehand can always be useful in order to be better prepared, locating clusters of utmost importance, and making policy decisions to contain the virus till the vaccines come out. A plethora of researchers worldwide are already trying to predict the spread of the virus in various ways \cite{Song2020.02.29.20029421, SIORDIA2020104357, giordano2020modelling}. In this paper, we focus on analyzing the compartmental epidemiological model with an emphasis on computing the basic reproductive number, commonly known as $R_0$. We focus on the following two aspects: (i) Inter-state migration within the country, (ii) The asymptomatic population. This is by no means meant to be an all-encompassing discussion of infectious disease modeling but a resource to supplement other more comprehensive texts are, \cite{Anderson1992}, \cite{Brauer2001},  \cite{Brauer2008} and \cite{Keeling2008}. The following section introduces the model that is developed in this paper.

\subsection{Introduction to the Model}

The SIR epidemic model is a widely used simplest compartmental model was introduced by Kermack and McKendrick (1927)~\cite{Kermack-McKendrick-SIR}. A compartmental model denotes mathematical modeling of infectious diseases where the population is separated in various compartments, S, I, or R, (Susceptible, Infectious, or Recovered). In this paper, we have developed an extension of SIR model, again a compartment based epidemiological model named SINTRUE to track and predict the spread of the virus and have demonstrated the model by applying it on the data available for the state of Chhattisgarh in India~\footnote{Obtained from Pt. J.N.M. Medical College Raipur, Chhattishgarh}. We have sourced our data from the official reports of The Government of India which is unofficially collated in an open source project called COVID-19 India Project \cite{covid19indiaorg2020tracker}. The model presented in this paper is developed as close to reality as possible, keeping in mind the different contagion rates from the asymptomatic (interchangeable with pre-symptomatic in this paper \footnote{We have interchangeably used asymptomatic and pre-symptomatic patients here, as our model allows us to put them in the same compartment as long as they do not show symptoms, and then assign to different compartments based on if they become symptomatic or remain asymptomatic. We understand that these two words are medically different, but the construction of the model allows us to use them interchangeably from an epidemiological perspective.}) and symptomatic patients, inflow to and outflow from the population due to migration, possibility of reinfection, and the extent of testing. The model is novel in the aspect of granularity, and specially in the aspect of considering a dynamic population that is often not considered in compartment based epidemiological models. This model is shown to be able to provide us with an estimate of the `unseen' COVID-19 infected patients in the population, and most importantly, the addition of a compartment of `unrecorded' recovery is able to provide a hint towards the status of herd immunity of the population. Also, this model, once applied on the data is able to show the presence of second wave of infection as well. This model has been shown to perform better than the other epidemiological models for COVID-19, and the results from this model has been able to provide concrete policy suggestions for healthcare management in the time of the pandemic.

The SINTRUE model comprises of seven compartments in the progression of the disease, with the addition of an inflow to and an outflow of people from the population. The seven compartments considered in the model are \textbf{S}usceptible, \textbf{I}nfected and pre-symptomatic, Infected and Symptomatic but \textbf{N}ot Tested, \textbf{T}ested Positive, Recorded \textbf{R}ecovered, \textbf{U}nrecorded Recovered, and \textbf{E}xpired.

The details of the model along with the description of the infection dynamics and the compartments are elaborated in Section \ref{model}. Section \ref{estim} details the estimation process of the parameters of the model. Finally, Section \ref{res} concludes with the results and discussions.

\section{The Model}\label{model}

\subsection{Developing the model}
To incorporate a realistic viewpoint of the dynamics of COVID-19 spread across India, we propose an extension of SIR type model, with $7$ compartmental states. There was interstate movements of migrant workers during the period of lockdown which was an undeniable part of the vital dynamics, was considered in only a few previous works \cite{MAJI2020100187}, but remained a concern for the media and the governments  \cite{indiatoday-migrant, joshi2020lockdown, migrant-issue-cabinet}. In our proposed model, we incorporate this effect of interstate migrant movement by keeping the total population dynamic, with an incoming population of migrants divided into three types: Not Infected ($M_n$), Pre-Symptomatic $(M_p)$, and Symptomatically Infected $(M_i)$, as well as an outgoing population of migrants from the set of population who are not under medical surveillance. The part of the incoming migrants who are not infected, pre-symptomatic or symptomatically infected join the corresponding part of susceptible $S$, pre-symptomatic $I_p$ and symptomatic and tested people $I_t$ compartments at destination. 
Starting from three compartments of SIR model\cite{Kermack-McKendrick-SIR}, we primarily  extend it by three aspects: 
\begin{enumerate}
    \item Adding another compartment $E$ concerning the recorded deaths of patients suffering from COVID-19.
    \item Splitting up the group of infectious persons into three compartments.
    \begin{enumerate}
        \item Pre-symptomatic individuals $I_p$, who do not show at least one of the primary symptoms of the disease. While in the literature this compartments is popularly known as asymptomatic patients \cite{asymptomatic-carrier}, we prefer to call this pre-symptomatic going by the analysis of World Health Organization(WHO) \cite{who-presymptomatic}, which clearly corroborates in favour of pre-symptomatic transmission rather than truly asymptomatic transmission.
        \item Symptomatic but not tested individuals namely $I_{sn}$. These are the people who despite being symptomatic is not under any kind of medical surveillance.
        \item Tested Positive individuals, $I_t$. This is the part of the population who are tested positive for COVID-19 and is under medical surveillance. Hence, the reported figures of COVID-19 patients only related to this compartment.
    \end{enumerate}
    \item Splitting up the group of recovered persons into two compartments.
    \begin{enumerate}
        \item $R$, the recorded part of the population recovered from COVID-19. 
        \item $U$, the unrecorded part of the population recovered from COVID-19. These people have gained immunity from the disease by recovering from it naturally, however, their situation is not reported to the hospitals, governments or any concerning agencies.
    \end{enumerate}
\end{enumerate}


\begin{figure}
    \centering
    \begin{tikzpicture}[->,>=stealth',auto,node distance=1cm,
  thick,main node/.style={circle,draw,font=\sffamily\Large\bfseries}]
  
        \node (S) [S_node] {$S$};
        \node (Ip) [Ip_node, right of = S, xshift = 2cm] {$I_p$};
        \node (It) [It_node, right of = Ip, xshift = 5cm] {$I_t$};
        \node (Isn) [It_node, below of = It, xshift = -3.5cm, yshift = -0.75cm] {$I_{sn}$};
        \node (U) [U_node, below of =Ip, xshift= 1cm, yshift=-2cm ]{$U$};
        \node (E) [E_node, below of = It, xshift = -2cm, yshift = -2cm]{$E$};
        \node (R) [R_node, below of = Isn, yshift = -2cm]{$R$};
        \node (In) [In_node, above of = S, xshift= 3cm, yshift=3.5cm] {In-Migrants};
        \node (Mn) [Mn_node, below of = In, xshift=-4cm, yshift=0cm] {Not Infected $(M_n)$};
        \node (Mp) [Mp_node, below of = In, xshift=0cm, yshift=-1cm] {Pre Symptomatic and Infected $(M_p)$};
        \node (Mi) [Mi_node, below of = In, xshift=5cm, yshift=-2cm] {Symptomatic and Infected $(M_i)$};
        \node (Out) [Out_node, below of=S, xshift=-2cm, yshift=-2cm] {Out-Migrants};
        
   
    \path[every node/.style={font=\sffamily\small}]
        (S) edge node [right, anchor = south] {\large $\beta_{1t}, \beta_{2t}$} (Ip)
        (Ip) edge node [right, anchor = south] {\large $\theta_t$} (It)
        (Ip) edge node [right, anchor = south] {\large $\alpha$} (Isn)
        (Ip) edge node [right, anchor = west] {\large $\lambda$} (U)
        (It) edge node [right, anchor = east] {\large $\tau$} (E)
        (It) edge[bend left = 40] node [right, anchor = west] {\large $\gamma$} (R)
        (R) edge[bend left = 40] node [left, anchor = south] {\large $f$} (S)
        (U) edge[bend left = 20] node [left, anchor = south] {\large $f$} (S)
        (Isn) edge node [right, anchor = south] {\large $\delta_t$} (It)
        (Isn) edge node [right, anchor = north] {\large $\zeta$} (E)
        (Isn) edge node [right, anchor = south] {\large $\kappa$} (U)
        (S) edge node [right, anchor = south] {} (Out)
        (Ip) edge node [right, anchor = south] {} (Out)
        (Mn) edge[bend right = 30] node [right, anchor = south] {} (S)
        (Mp) edge node [right, anchor = south] {} (Ip)
        (Mi) edge node [right, anchor = south] {} (It)
        (In) edge node [right, anchor = south] {\large $c_1$} (Mn)
        (In) edge node [right, anchor = east] {\large $c_2$} (Mp)
        (In) edge[bend left = 30] node [right, anchor = south] {\large $c_3$} (Mi);
    \end{tikzpicture}
    \caption{Interaction graph between different states of the model}
    \label{fig:interaction-graph}
\end{figure}
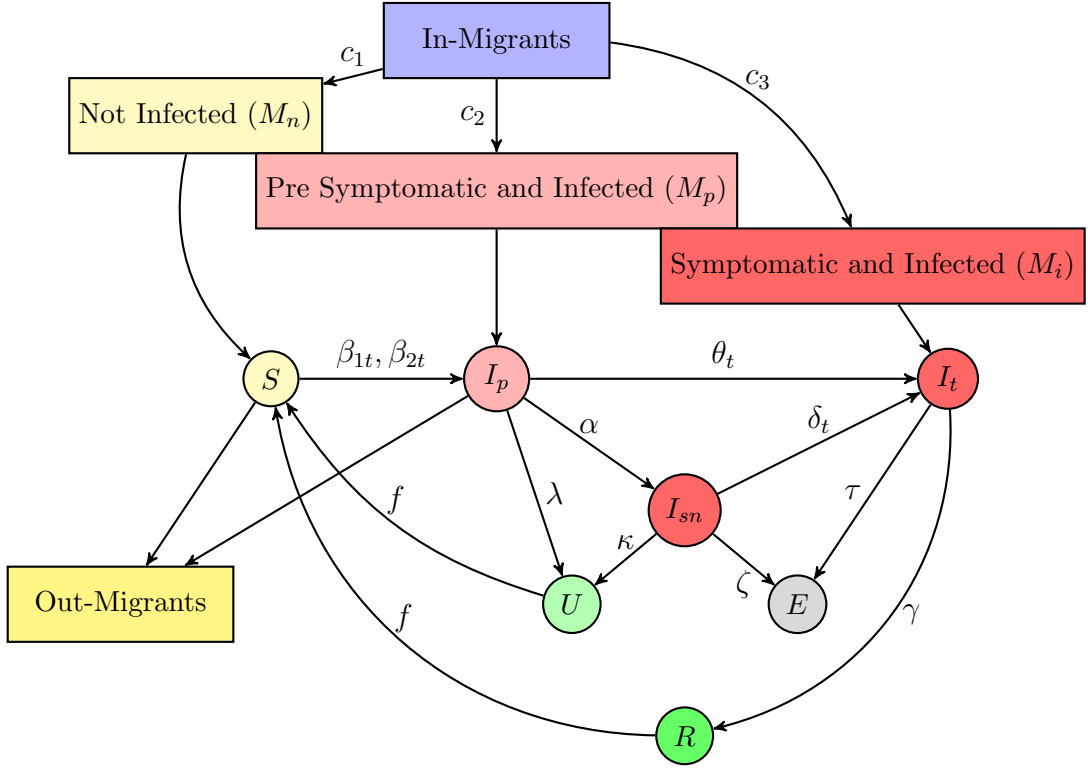
The model is visually explained in~\cref{fig:interaction-graph} along with the corresponding notations. Starting with the susceptible population, a person might get corona-virus from his/her interaction with a asymptomatic person, as well as a symptomatic person. While there is a natural predilection to avoid symptomatic persons, such a tendency is not prevalent in interaction with asymptomatic persons due to lack of evidence through external appearance. 
Also, there is evidence of difference in transmission for separate viral loads, complex incubation periods and rates of disease progression~\cite{tan-j-asymptomatic}. Furthermore, as awareness about the disease begin to spread, the general population tend to avoid too many interactions, as well as reduce the exposure time periods. Thus, we take $\beta_{1t}$ and $\beta_{2t}$ as the time varying transmission rates, possibly different, for the interaction between the susceptible population with the pre-symptomatic $I_p$ and symptomatic not tested $I_{sn}$ population. The implicit assumption that the transmission rate between susceptible and symptomatic tested population is $0$, can be justified by the fact that the people under medical surveillance is being put under quarantine, restricted to interact with susceptible population, and health workers interact with these patients with proper Personal Protective Equipment (PPE) so that the rate of transmissions through these interactions is negligible.

In addition to these disease transmissions, the transitions between different compartments of infectious states in our model is based on a single principle. At any point of time, an infectious person is either recovered naturally, or the disease aggravates to next stage, or the person comes under medical surveillance, by means of contact tracing. While there is little evidence of evolution of corona-virus during the time span of the pandemic~\cite{maclean-covid-evolution}, there were significant changes in testing strategies~\cite{icmr-test-strategy-v2, icmr-test-strategy-v3, icmr-test-strategy-v4, icmr-test-strategy-v5}. Thus, it is reasonable to assume that the rate concerning with exacerbation of the disease or natural cure of the disease are not time varying as they are related to inherent biological variables immune to external circumstantial changes, in contrast, the rates concerning the inclusion of population under medical surveillance should obviously remain a time varying quantity. Under this logical flow, a pre-symptomatic person in $I_p$, naturally heals and moves to $U$ following a counting process with rate $\lambda$, or starts to show symptoms (next stage of the disease) following another independent counting process with rate $\alpha$, or comes under medical surveillance following a time varying rate $\theta_t$. Exactly similar to that, a symptomatic but not tested person $I_{sn}$ naturally heals and moves to $U$ with rate $\kappa$, or the disease proceeds to next stage resulting in death of the person with rate $\zeta$, or gets tested positive with rate $\delta_t$.

Finally, similar to SIR model, the people under medical surveillance $I_t$, either recovers with rate $\gamma$ or becomes deceased with rate $\tau$. We also allow for the possibility of re-infection, by assuming that a proportion $f$ of the people getting recovered each day, will actually join the susceptible group instead, due to lack of sufficient antibodies. The complete dynamics of the model can be mathematically expressed using a set of differential equations as shown in~\cref{eqn:diff}\footnote{Note: This model does not have mass conservation property as we are considering a dynamic population whose input and output rates are not equal.}.

\begin{equation}
    \begin{split}
    \diff{N_t} & = -\diff{\text{(Out)}} + \diff{\text{(In)}} \\
    \diff{\text{(In)}} & = \diff{M_n} + \diff{M_p} + \diff{M_i}\\
    \diff{M_n} & = c_1 \diff{\text{(In)}}\\
    \diff{M_p} & = c_2 \diff{\text{(In)}}\\
    \diff{M_i} & = c_3 \diff{\text{(In)}}\\
    \diff{S} & = -\beta_{1t} \dfrac{SI_p}{N_t} - \beta_{2t} \dfrac{SI_{sn}}{N_t} + f\diff{(R + U)} - \diff{\text{Out}} + \diff{M_n} \\
    \diff{I_p} & = \beta_{1t} \dfrac{SI_p}{N_t} + \beta_{2t} \dfrac{SI_{sn}}{N_t} + \diff{M_p} - \alpha I_{p} - \theta_t I_p - \lambda I_p \\
    \diff{I_{sn}} & = \alpha I_p - \delta_t I_{sn} - \kappa I_{sn} - \zeta I_{sn} \\
    \diff{I_t} & = \theta_{t} I_p + \delta_t I_{sn} + \diff{M_i} - \gamma I_t - \tau I_t\\
    \diff{R} &  = \gamma I_{t} - f \diff{R}\\
    \diff{E} & = \tau I_t + \zeta I_{sn}\\
    \diff{U} & = \lambda I_p + \kappa I_{sn} - f \diff{U}\\
    \label{eqn:diff}
    \end{split}
\end{equation}

\subsection{Simplifying assumptions}

While the general model described in \cref{eqn:diff} is more close to the reality, it is also very complex to analyze. In light of the available data, certain assumptions are needed in order to effectively estimate all the parameters of the model. 

The first implicit assumption in the model dynamics is that interactions between different states of the model can be efficiently modelled by a counting process with the respective rate parameters. In particular, if we have an arrow from state $A$ to state $B$ with rate $\theta$ in \cref{fig:interaction-graph}, then it means the number of people that moves from state $A$ to state $B$ at any given day is determined by a Poisson process with parameter $\theta$, in other words, an individual residing in state $A$ moves to state $B$ after an exponentially distributed number of days, with mean parameter $1/\theta$.

One crucial assumption that we need to make is that $f = 0$. This is actually supported by dearth of strong evidences of reinfection of any recovered patients, with very inconsistent and sporadic incidents\cite{reinfection-alizargar, who-reinfection-news,roy2020covid}. Next, we take $\zeta = 0$, with the implicit assumption that any deaths due to COVID-19 will be reported. Also, as we do not have any distinction in the deceased data about the sources of the deceased (whether they were monitored under medical surveillance prior to death), we are compelled to make this assumption in order to ensure estimability of the parameters presented in our model.

In addition to this, we also assume that the transmission rates $\beta_{1t}$ and $\beta_{2t}$ differs only by the different exposure time with the susceptible population. Since the type of coronavirus associated with COVID-19 is relatively new, no significant biological study has yet measured the contagiousness of the virus transmitting from a pre-symptomatic individual apart from a symptomatic individual. In this regard, a susceptible person in the population can differentiate between a pre-symptomatic and a symptomatic stranger only on the basis of symptoms that are externally visible like cough, but not on the basis of symptoms like fever, fatigue, increased blood pressure, low oxygen levels etc. as they are improbable to perceive from external appearance without any assist of proper medical instruments. Thus, we may assume $\beta_{2t} = P(\text{Cough})\beta_{1t}$ for any $t$. According to \cite{SIORDIA2020104357}, the rate of symptoms like cough shown along with positive COVID-19 cases is $61.7\%$, and thus we take $\beta_{2t} = 0.617\beta_{1t}$ as a restrictive assumption in our model.

In order to estimate the transmission rate $\beta_{1t}$, we shall assume a specific two parameter family of curves. 

\begin{equation}
    \beta_{1t}(s) = \begin{cases}
    a + e^{-bs} & \text{ if } s \text{ is before 1st May, 2020}\\
    a & \text{ otherwise }
    \end{cases}
    \label{eqn:beta1-parametrization}
\end{equation}

where $a, b$ are parameters to be estimated. The threshold of 1st May, 2020 is carefully chosen to distinguish the period that denotes the beginning of official migrant movement~\cite{hindu-shramik-special-train}. Before this official migration started, various lockdown enforcement schemes were expected to reduce the transmission rate continuously. However after 1st May of 2020, various psychological issues among migrants~\cite{migrant-psychology-report}, the difficulty of lockdown enforcements~\cite{impossible-social-distancing, covid-oxymoron}, aggressive violence against health care workers~\cite{black-day-covid} suggests that transmission rate cannot be reduced indefinitely and should remain constant in a state of partial lockdown.

\section{Estimation of the Model Parameters}\label{estim}

We start by delineating the estimation procedure for parameters associated with migrant movement, i.e $c_1$, $c_2$ and $c_3$. The total number of incoming and outgoing migrants can be estimated as in~\cite{MAJI2020100187}. 

Note that $c_3$ is essentially the proportion  of \textit{incoming} migrants who are symptomatic and infected, i.e they have contracted the virus in the origin state and have developed the symptoms by the time the reached the destination state. We use \textit{Destination State} to denote the particular administrative area serving as the destination of the in-migrants. Similarly, \textit{Origin State} is used to denote the administrative area from which an in-migrants starts his journey to the Destination state. Let $D$ denote a destination state and $O_1, \cdots, O_n$ denote $n$ origin states. 

Let 
\begin{align*}
  \Prob\left(O_i \ \text{is the Origin state }|\  D \ \text{is the destination state}\right) &= m_i \quad, \quad i=1(1)n\\  
\Prob\left(\text{Active cases of Out-Migrants in state $O_i$} \right)&= A_i \quad, \quad i=1(1)n\\  
\end{align*}
Let $m_D=(m_i)_{i=1(1)n}$ and $A_D=(A_i)_{i=1(1)n}$. Then we estimate $c_3$ by estimated proportion of reported Covid-19-positive people among incoming migrants in each day. Thus,
\[ \widehat{c_3}=\displaystyle \sum_{i=1}^n m_i A_i=m_D^TA_D  \]

We do not yet have a origin-state-specific segregation vector of Migrants entering a destination state each day- normalizing which would have yielded $m_D$ . But we have a data\footnote{Obtained from Pt. J.N.M. Medical College Raipur, Chhattishgarh} , on what proportion of \textit{Infected Incoming Migrants} are from which Origin state. We use that vector as $\widehat{m_D}$. This is because, considering the relatively large number of total migrant influx compared to \textit{infected} migrant influx (i.e belonging to state $I_t$), and more or less large number of infections in all the significant origin states, we can assume that the infection among in-migrants in independent of the origin state. Mathematically this assumption can be written as
\[ \Prob\left(\text{ Origin state }= O_i | \text{ Infected in-migrant }\right)=\Prob\left(\text{ Origin state }= O_i \right| \text{ In-migrant }) \]
This assumption validates $\widehat{m_D}$ as an estimate of $m_D$.
\par

We estimate $A_D$ by assuming that the \textit{Other States} and/or \textit{Unassigned/Unknown} column in the data provided by Covid19India dashboard~\cite{covid19indiaorg2020tracker} corresponds to migrant people. We have an estimate $M_i, \ i=1(1)n$ for the Total number of migrants likely to move, as in~\cite{MAJI2020100187}. Let at time point $t$, the total number of Unassigned cases in Origin state $O_i$ be denoted as $\left(\mathcal{A}_i\right)_t$.

Let $A_t=\left(\dfrac{\left(\mathcal{A}_1\right)_t}{M_1}, \ \dfrac{\left(\mathcal{A}_2\right)_t}{M_2}, \cdots, \ \dfrac{\left(\mathcal{A}_n\right)_t}{M_n}\right)$. 
Then, $c_3$ at time point $t$ is estimated as:
\[ \widehat{c_3}_t=\widehat{m_D}^TA_t \]
To compute $c_2$ we note that those incoming migrants who  are undetected would have contracted the disease during the journey with fellow migrants. Thus  at time point $t$, a symptomatic migrant from origin state  $O_i$ would infect on an average $(R_i)_t$ people, where $R_i$ denote the day-wise reproduction number of the $i$-th origin state. We use \texttt{EpiEstim} package\cite{coriEpiEstim} to obtain these reproduction numbers.

Let $R_t=\left((R_1)_t,, \cdots  , (R_n)_t \right)^T$. Then $c_2$ at time point $t$ is estimated as:
\[ \widehat{c_2}_t=\left(\widehat{m_D} \odot A_t\right)^T R_t \]
where $\odot$ denote the Hadamard Product i.e element-wise multiplication. Finally, we obtain;

\begin{equation*}
    \widehat{c_1}=1-\widehat{c_2}-\widehat{c_3}
\end{equation*}


Now we move on to estimating other parameters. Once we assume that $f = 0$ and $\zeta = 0$, the differential equations described the changes of the states $E$ and $R$ simplifies to;

\begin{equation}
    \begin{split}
        \diff{R} & = \gamma I_t\\
        \diff{E} & = \tau I_t\\
    \end{split}
    \label{eqn:diff-RE-simple}
\end{equation}

Since the periodically released data by Indian government~\cite{govt-covid-dashboard}, along with their historical records by Covid19India group~\cite{covid19indiaorg2020tracker} contains records of recovered, deceased and current number of hospitalized covid patients, the quantities $\diff{R}, \diff{E}$ and $I_t$ can be computed with the help of the data and standard numerical differentiation techniques. Now, we choose an $L_1$ norm of the errors as a criterion for estimating $\gamma$ and $\tau$, i.e.

\begin{equation}
    \hat{\gamma} = \sum_i \left\vert \left(\diff{R}\right)_i - \gamma (I_t)_i \right\vert \quad \text{and} \quad \hat{\tau} = \sum_i \left\vert \left(\diff{E}\right)_i - \tau (I_t)_i \right\vert
    \label{eqn:tau-gamma-est}
\end{equation}

which essentially boils down to find robust estimators for MAD (median absolute deviation) regression problems to express the dependent variable $\diff{R}$ and $\diff{E}$ as a intercept-free linear function of independent variable $I_t$. 

From the testing strategy mentioned in ~\cite{icmr-test-strategy-v5}, it is evident that pre-symptomatic people are getting tested only if they are traced as high-risk contacts of a confirmed case. In other words, a pre-symptomatic person would be tested if any person that he / she had contacted for last few days has coronavirus and is tested positive. Now, turning our attention to a single asymptomatic person, let $N$ be a random variable denoting the number of contacts that the person had in past few days. Assume a Poisson distribution for $N$ with mean parameter $N_c$. Now note that, 
    
$$\text{Number of symptomatic contacts}\mid N \sim Bin\left( N, p_s \right)$$
    
and similarly,
    
$$\text{Number of asymptomatic contacts}\mid N \sim Bin\left( N, p_a \right)$$
    
where $p_s$ and $p_a$ denotes these probabilities for an individuals to be symptomatic or pre-symptomatic given that he/she is not already tested. Note that both of these random variables are independent of each other and also the unconditional distribution of number of symptomatic (or asymptomatic) contacts is Poisson distributed with mean parameter given by product of $N_c$ and the corresponding binomial success probability. Thus,
    
\begin{align*}
    e^{-\theta_t d}
    & = \Prob(\text{The person gets tested in } \geq d \text{ days})\\
    & = \Prob(\text{all symptomatic and asymptomatic contacts gets tested in } \geq d \text{ days})\\
    & = \left( \sum_{n = 0}^{\infty} e^{-n\theta_t d} e^{-N_c p_a} \dfrac{(N_c p_a)^n}{n!} \right) \times \left( \sum_{n = 0}^{\infty} e^{-n\delta_t d} e^{-N_c p_s} \dfrac{(N_c p_s)^n}{n!} \right)\\
    & = \exp\left[ -N_c p_a (1 - e^{-\theta_t d}) -N_c p_s (1 - e^{-\delta_t d})\right]\\
    & \approx \exp\left[ -(N_c p_a \theta_t +N_c p_s \delta_t) d\right], \qquad \text{assuming the respective quantities are small}
\end{align*}

which finally implies; $\theta_t \approx N_c(p_a \theta_t + p_s\delta_t)$. To obtain the probabilities $p_a$ and $p_s$, it follows from the model that a plug-in estimator would be $I_p / (N_t - I_t - E - R)$ and $I_{sn}/(N_t - I_t - E - R)$, which is basically the ratio of the corresponding pre-symptomatic (or symptomatic) individuals in the population and the individuals not under any kind of medical surveillance (i.e. they are at risk of getting tested). However, since $(I_p \theta_t + I_{sn} \delta_t)$ is the total number of individuals getting tested on day $t$ (see \cref{eqn:diff}), the estimating equation for $\theta_t$ simplifies to;

\begin{equation}
    \theta_t \approx \dfrac{N_c}{(N_t - I_t - E - R)} \Delta(I_t + E + R)
    \label{eqn:theta-est}
\end{equation}

A quick interpretation of the above estimating equation is that, when $\Delta(I_t + E + R)$ new individuals are tested positive on day $t$, by means of contact tracing, on average $N_c \times \Delta(I_t + E + R)$ many individuals will be tested for the disease next day. Thus, an pre-symptomatic individual, who is very similar to a healthy individual in external factors would have the same rate of being selected at the testing as the ratio of $N_c \times \Delta(I_t + E + R)$ and remaining population $(N_t - I_t - E - R)$. Since $N_c$ is an unknown parameter, we take it as the average number of high risk contacts traced (and tested for covid) per covid patients who are tested positive. 

Focusing on the testing rate $\delta_t$ for the symptomatic individuals, we again use the assumption that the diffusion from $I_{sn}$ occurs according to independent Poisson processes with corresponding rates, and thus at any point of time $T$, the probability that a symptomatic individual will get tested before he/she recovers (or dies) is; $$\left(\int_0^T \delta_t(s)ds / \int_0^T (\delta_t(s) + \kappa + \zeta)ds\right)$$
On the other hand, a simple application of Bayes Rule enables us to write the probability as;

\begin{equation}
    \begin{split}
        & \Prob(\text{Tested} \mid \text{Symptomatic and has covid})\\
        = \quad & \dfrac{\Prob(\text{Tested and Symptomatic and has covid})}{\Prob(\text{Symptomatic and has covid})} \\
        = \quad & \dfrac{\Prob(\text{Symptomatic} \mid \text{Tested and has covid}) \Prob(\text{Tested and has covid}) }{\Prob(\text{Symptomatic and has covid})} \\
        = \quad & \dfrac{\Prob(\text{Symptomatic} \mid \text{Tested and has covid}) \Prob(\text{Tested and has covid}) }{\Prob(\text{Symptomatic} \mid \text{Covid}) \Prob(\text{has covid})} \\
    \end{split}
    \label{eqn:bayes-symptomatic}
\end{equation}

From the patient database available from Chattishgarh government, it was possible to obtain the number of symptomatic people among those who are tested positive for COVID-19, thus enabling us to obtain estimate of $\Prob(\text{Symptomatic} \mid \text{Tested and has covid})$. Also, the quantity $\Prob(\text{Tested and has covid})$ can be estimated from the number of samples tested for COVID-19. Note that, the estimates for the quantities in the denominator cannot be obtained in the context of the particular country in study, since only the data pertaining to the tested individuals will be collected. However, Siordia~\cite{SIORDIA2020104357} reported the rate of different symptoms seen along with COVID-19 patients, which can be used to estimate $\Prob(\text{Symptomatic} \mid \text{Covid})$ based on the definition of symptomatic person as an infected individual showing all of the primary 3 symptoms, namely fever, cough and fatigue. This turns out to be, $\Prob(\text{Symptomatic} \mid \text{Covid}) = (0.822 \times 0.617 \times 0.44) = 0.22315656$.

To estimate the true prevalence $\Prob(\text{has covid})$, we consider to use the data of countries like South Korea, United Arab Emirates as a reference frame, since these countries perform most of the tests in relative to their population~\cite{owidcoronavirus}. The estimates of test positivity rates for these countries ranges between $2.5\%$ to $5\%$ on average. Some studies~\cite{10.1093/cid/ciaa761} shows that the prevalence of COVID-19 among the health workers who are high-risk contacts stand somewhere close to $5\%$. Turning our attention to the available data, the prevalence among the high risk contacts who are tested stand at $4.229\%$ for Chattishgarh and $8.642\%$ for the whole India. Naturally, it is apparent that the true prevalence would be lower than $4.229\%$, but it needs to be estimated in a proper way. Let us denote this true prevalence as $\epsilon$, which we shall also estimate in the light of available data.

Assuming the knowledge of $\epsilon$ and $\kappa$, we can obtain a plug-in estimate for $\delta_t$, as $\zeta = 0$ by assumption and

\begin{equation}
    \delta_t(s) = \dfrac{d}{ds} \left[\kappa s  \dfrac{(1 - \Prob(\text{Tested} \mid \text{Symptomatic and has covid})(s) )}{\Prob(\text{Tested} \mid \text{Symptomatic and has covid})(s) }\right]
    \label{eqn:estimate-delta}
\end{equation}

where the probability function can be evaluated using \cref{eqn:bayes-symptomatic} using the knowledge of the true prevalence $\epsilon$ at the discrete timepoints, and then a numerical differentiation can be performed to obtain the time-varying rate $\delta_t$.

Turning our attention to estimation of the transmission rate, due to \cref{eqn:beta1-parametrization}, it follows that the knowledge of $a, b$ will enable one to estimate $\beta_{1t}$ and thus in turn $\beta_{2t}$ as $0.617\beta_{1t}$. Thus with a specific choice of $\alpha, \lambda, \kappa, \epsilon, a, b$, it is possible to obtain the time varying parameters $\beta_{1t}, \beta_{2t}, \delta_t$. Combining these with the already obtained estimates $\theta_t, \tau, \gamma$ and time varying estimates regarding migrant movements $c_1, c_2, c_3$, it is possible to simulate the whole system by solving the set of differential equations \cref{eqn:diff} numerically. Denoting $\phi = \left( \alpha, \lambda, \kappa, \epsilon, a, b \right)$, in order to obtain an estimate of the parameter $\phi \in \Phi \subseteq \mathbb{R}^6$ where $\Phi$ denotes the underlying parameter space, we consider the following criterion.

\begin{equation}
    \mathcal{L}(\phi) = w_{I} \mathcal{L}_I(\phi) + w_R \mathcal{L}_R(\phi) + w_E \mathcal{L}_E(\phi) + w_{\text{symp}} \mathcal{L}_{\text{symp}}(\phi) + w_{\text{asymp}} \mathcal{L}_{\text{asymp}}(\phi) + w_{\text{inf}} \mathcal{L}_{\text{inf}}(\phi)
    \label{eqn:LS-phi}
\end{equation}

where $w$'s are some specifically chosen weights and;

\begin{align*}
    \mathcal{L}_I(\phi) & = \sum_{s = 1}^{T} (I_t^{\text{data}}(s) - I_t(s)(\phi))^2\\
    \mathcal{L}_R(\phi) & = \sum_{s = 1}^{T} (R^{\text{data}}(s) - R(s)(\phi))^2\\
    \mathcal{L}_E(\phi) & = \sum_{s = 1}^{T} (E^{\text{data}}(s) - E(s)(\phi))^2\\
    \mathcal{L}_{\text{symp}}(\phi) & = \sum_{s = 1}^{T} (\Delta I^{\text{data}}_{t, \text{symp}}(s) - \widehat{\theta_t}(s)I_p(s)(\phi))^2\\
    \mathcal{L}_{\text{asymp}}(\phi) & = \sum_{s = 1}^{T} (\Delta I^{\text{data}}_{t, \text{asymp}}(s) - \widehat{\delta_t}(s)I_{sn}(s)(\phi))^2\\
    \mathcal{L}_{\text{inf}}(\phi) & = ((1 - \epsilon)N_t(T)(\phi) -  S(T)(\phi))^2\\
\end{align*}

The above loss functions measure difference in the the simulated counts and the real counts from the available data. For instance, $\mathcal{L}_I(\phi), \mathcal{L}_R(\phi), \mathcal{L}_E(\phi)$ measures the discrepancy between the available data of the number of infected, recovered, deceased individuals and the number of infected, recovered, deceased individuals obtained from the numerical solution to \cref{eqn:diff} with the choice of the parameters $\phi$. On the other hand, $\mathcal{L}_{\text{symp}}(\phi)$ and  $\mathcal{L}_{\text{asymp}}(\phi)$ measures the discrepancy between the obtained data on the number of new symptomatic and pre-symptomatic patients tested positive on each day, with that of the theoretical counterparts $\theta_t I_p$ and $\delta_t I_{sn}$ respectively. Finally, at the most recent timepoint $T$ with available data, $(1 - \epsilon)N_t$, which denotes the number of individuals without COVID-19 virus (since $\epsilon$ is the true prevalence of COVID-19), is matched against the number of susceptible population $S$. The best fitting parameters $\widehat{\phi}$ is obtained by minimizing $\mathcal{L}(\phi)$. However, in order to circumvent numerical underflow or overflow and convergence related issues due to the possibility of non-convex optimization with numerous local minima, we decided to use grid search algorithm to find the best $\phi$ which minimizes $\log \mathcal{L}(\phi)$. The weights in \cref{eqn:LS-phi} are chosen carefully by performing a cross validation to minimize prediction sum of squares, in order to normalize each of the individual loss functions in the same range to make them comparable, so that all the loss functions are minimized simultaneously when minimizing $\log \mathcal{L}(\phi)$.

\section{Results \& Discussion}\label{res}

All the subsequent results are based on the available data as of August 22, 2020 collected from different official\cite{govt-covid-dashboard} and unofficial sources\cite{covid19indiaorg2020tracker}.

\begin{figure}[ht]
    \centering
    \includegraphics[width = \linewidth]{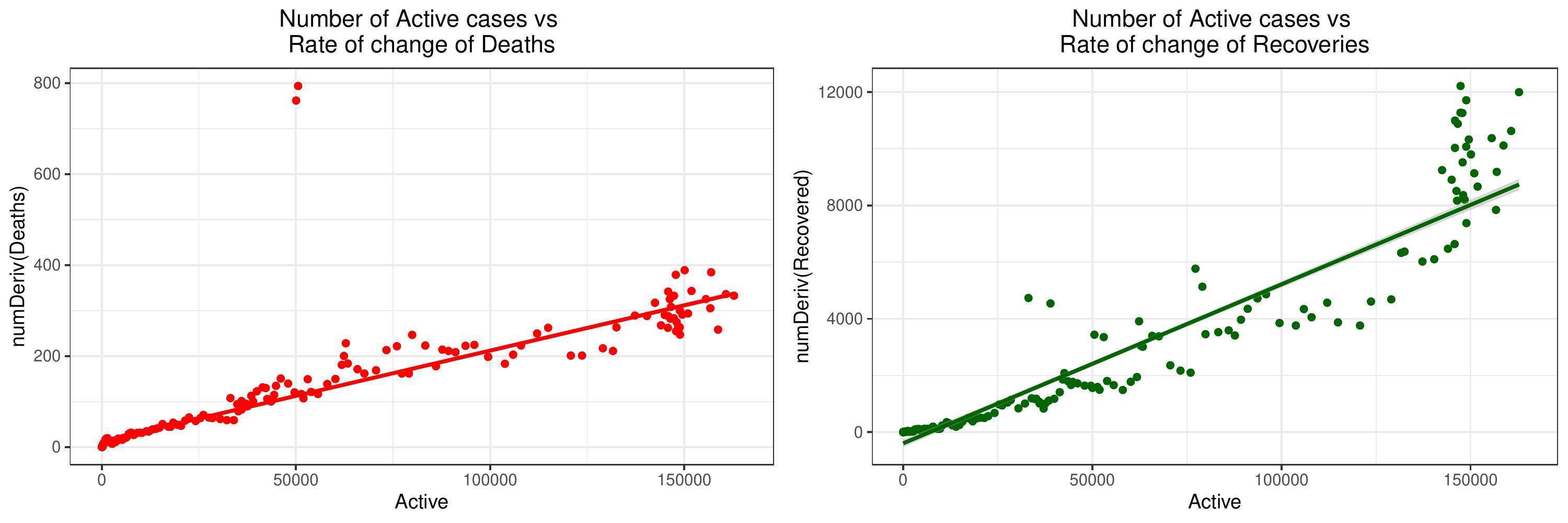}
    \caption{Goodness of fit for regression based approach to estimation of fatality and recovery rate for Maharashtra}
    \label{fig:Maharashtra-tau-gamma}
\end{figure}

Based on the robust regression type approach to estimate fatality rate $\tau$ and recovery rate $\gamma$, we use \texttt{MASS} package\cite{MASS-package} in \texttt{R} programming language to obtain these estimates. From the historical data on confirmed, active, recovered and deceased cases for each state, available at \cite{covid19indiaorg2020tracker}, we obtain the estimates of $\tau$ and $\gamma$ for some of the major states in India. The estimates are shown in \cref{tab:tau-gamma-est}. The fit of the robust regression type approach is shown in \cref{fig:Maharashtra-tau-gamma}.

\begin{table}[ht]
    \centering
    \begin{tabular}{|c|c|c|}
        \hline
         \textbf{State} & \textbf{Fatality Rate} ($\widehat{\tau}$)  &  \textbf{Recovery Rate} ($\widehat{\gamma}$)\\
         \hline
         Maharashtra & 0.00211257 & 0.05237954\\
         Tamil Nadu & 0.001665098 & 0.0937174\\
         Andhra Pradesh & 0.001014294  & 0.09728257 \\
         Karnataka & 0.00138392 & 0.06022071\\
         Delhi & 0.002318402 & 0.09441121\\
         Uttar Pradesh & 0.00135086 & 0.07263133 \\
         West Bengal & 0.002109226 & 0.09711365\\
         Gujarat & 0.002160267 & 0.06682686\\
         Chhattisgarh & 0.001200815 & 0.07298542\\
         \hline
    \end{tabular}
    \caption{Estimate of fatality rate and recovery rate for different states of India}
    \label{tab:tau-gamma-est}
\end{table}

It seems that \cref{fig:Maharashtra-tau-gamma} shows a reasonably good linear relationship between the numerical derivative of deceased and recovered cases, and the number of active cases, except a few outlying datapoints, which also suggests one possible reason for using a robust version of linear regression.

\begin{figure}[ht]
    \centering
    \includegraphics[width = 0.75\linewidth]{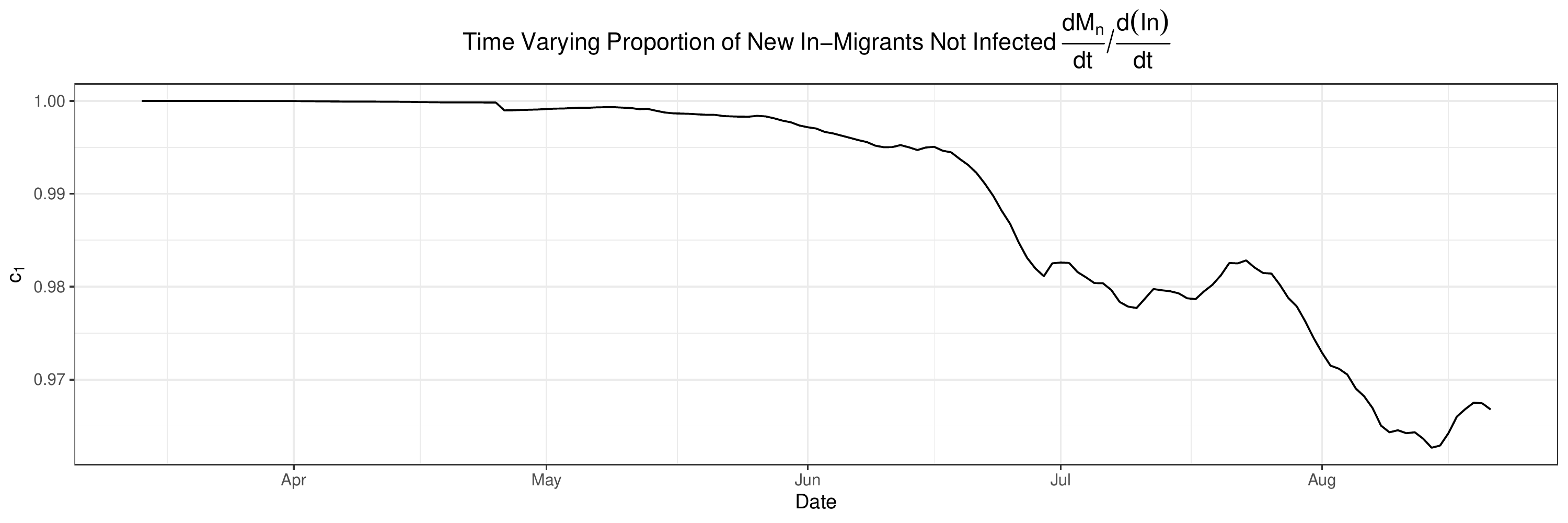}
    \includegraphics[width = 0.75\linewidth]{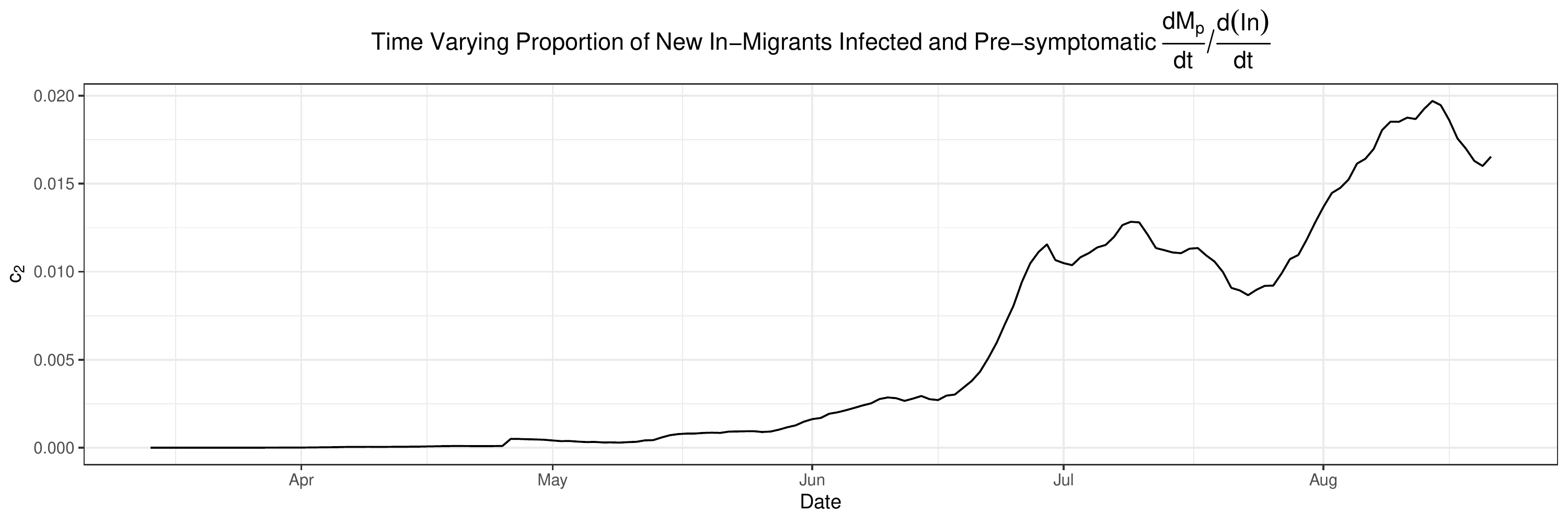}
    \includegraphics[width = 0.75\linewidth]{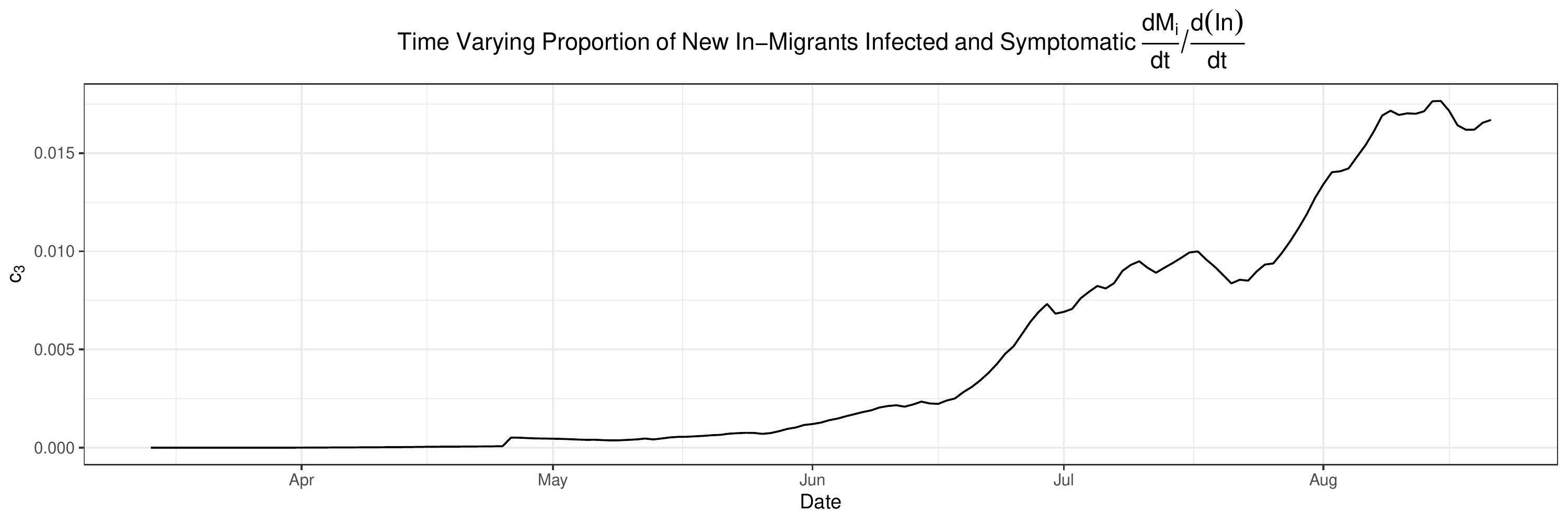}
    \caption{Estimated parameters pertinent to migration movement in Chhattisgarh}
    \label{fig:chhattisgarh-migration}
\end{figure}

Considering the migrant population, the total number of incoming and outgoing migrants can be estimated as in \cite{MAJI2020100187}. In case of Chhattishgarh, it was found that about $49.36\%$ of migrants come from Maharashtra, while migrants from Uttar Pradesh, Delhi, Telangana, Gujarat, Tamil Nadu, Haryana, Odisha, Andhra Pradesh constitutes more than $90\%$ of the incoming migrants. Thus when performing the estimation of parameters related to migrant movement, the reproduction number of these states, the number of special trains that connect Chhattisgarh to any of these states, the worker's population at the major cities of those states are taken into consideration. Based on the detailed estimation process described before, the time varying proportion of new incoming migrants based on different groups i.e. $c_1, c_2, c_3$ are estimated. Corresponding results for Chhattisgarh in shown in \cref{fig:chhattisgarh-migration}. The results show an overall decreasing pattern in $c_1$, while similar increasing pattern in $c_2, c_3$. Interestingly, a week before official migration started, the proportions $c_1, c_2$ becomes non-negative, which could be a possible indication of unofficial migrant movement. A sharp increase in $c_2$ and $c_3$ is also noticeable from mid-June, which was possibly a lagged effect of ``Unlock-1.0" declared by Govt. of India\cite{sharma2020unlock1}.

\begin{figure}[ht]
    \centering
    \includegraphics[width = \linewidth]{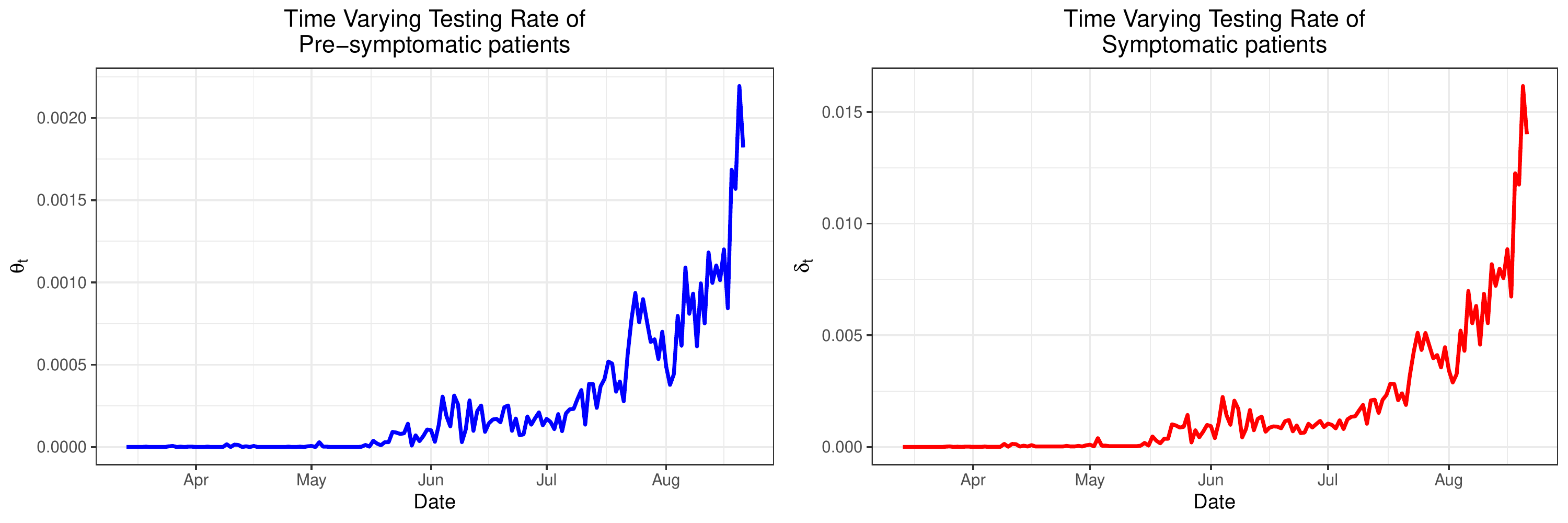}
    \caption{Time varying tested rate for pre-symptomatic and symptomatic individuals}
    \label{fig:chhattisgarh-theta-delta}
\end{figure}

Now turning our attention to the estimation of $\theta_t$ based on \cref{eqn:theta-est}, an important component to estimation of $\theta_t$ is the choice of $N_c$, the average number of contacts per day. Previous studies such as Mossong et al.\cite{Mossong2008} show that under natural circumstances, the average number of contacts of a person per day is $13.4$ with a varying distribution of mean number of contacts in different countries and in different regions with different socio-economic strata. Because of the skewness in the distribution of number of contacts, national lockdown restrictions and overall awareness about the disease spread, it is natural to assume that the average number of contacts is smaller than $13.4$. However, a more recent and relevant study by Leung et al.\cite{Leung2017} shows that average number of reported contacts in relevance to spread of a respiratory illness is much lower, ranges from $5.12$ to $8.21$ over different age groups and socio-economic stratum, with overall mean being $6.93$. In addition, from the official contact tracing data of Chhattisgarh, we have obtained that there were $23007$ primary contacts who are traced from $3257$ positive cases of covid patients, thereby suggesting an estimate of $N_c$ as $23007 / 3257 \approx 7.0638$. Since this matches nearly with the conclusions presented in \cite{Leung2017}, we take $N_c = 7.0638$ in our model and estimate $\theta_t$ using \cref{eqn:theta-est}. Similarly, minimizing \cref{eqn:LS-phi} enables us to find the optimized parameter $\kappa$, which in turn can be used to obtain the time varying testing rate for symptomatic individuals using \cref{eqn:estimate-delta}. The obtained estimated are shown in \cref{fig:chhattisgarh-theta-delta}. Although the estimated rates are highly correlated and shows similar patterns, the different in the magnitude depicts the fact that ceteris paribus, tracing a symptomatic individual and probability of him/her getting tested is about $7-8$ times as high as the same for a pre-symptomatic individual. However, while about $85-90\%$ of the new patients who are tested positive in Chhattisgarh are pre-symptomatic or mildly symptomatic, and as $\theta_t$ is estimated to be very small, it must be the case that the pool of pre-symptomatic individuals $I_p$ must be enormously high.

\begin{figure}
    \centering
    \includegraphics[width = \linewidth]{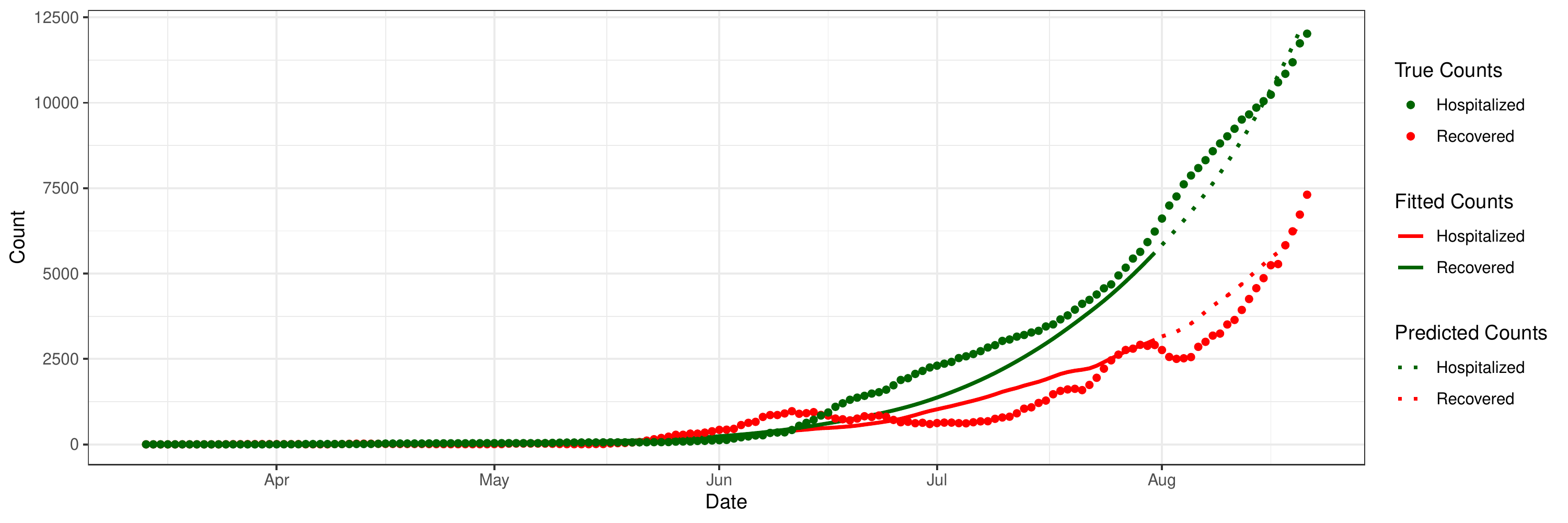}
    \caption{Fitted, True counts and Short term predicted counts of the number of hospitalized ($I_t$) cases and recovered cases $(R)$ in Chhattisgarh}
    \label{fig:chhattisgarh-fitted}
\end{figure}

\begin{table}[ht]
    \centering
    \begin{tabular}{|c|c|p{10cm}|}
         \hline
         \textbf{Parameter} & \textbf{Estimate} & \textbf{Explanation} \\
         \hline
         $\kappa$ & 0.0113 & On average, about $1.13\%$ of all symptomatic individuals recover from COVID-19 naturally every day. Compared to that, $1.5\%$ of all symptomatic individuals actually get tested positive every day as seen in \cref{fig:chhattisgarh-theta-delta}. \\
         \hline
         $\alpha$ & 0.012 & Different studies and reports\cite{who-presymptomatic} shows that average incubation period is about $4-5$ days. This means about $4.8\%$ to $6\%$ of newly infected pre-symptomatic individuals are expected to develop symptoms, before getting tested or naturally recovered.\\
         \hline
         $\lambda$ & 0.079 & About $7.9\%$ of the newly infected pre-symptomatic (or mildly symptomatic) individuals are expected to naturally recover from the disease.\\
         \hline
         $a$ & 0.083 & Tranmission rates decreased by a significant rate during the lockdown enforcement and before official permit of migration movements.\\
         \hline
         $b$ & 0.102 & Current transmission rate shows only $10.2\%$ of all interactions with pre-symptomatic individuals could expose the susceptible individual to the virus.\\
         \hline
         $\tau$ & 0.0012 & On average, about $0.12\%$ of the hospitalized cases turn to fatality every day.\\
         \hline
         $\gamma$ & 0.0729 & On average, about $7.29\%$ of the hospitalized cases turn to recoveries every day.\\
         \hline
         $\epsilon$ & 0.1955 & The true prevalence of COVID-19 is approximately $1.95\%$, thus affecting more than $6$ lakhs individuals among the $3.2$ crore population of Chhattisgarh. \\
         \hline
    \end{tabular}
    \caption{Estimates of the parameters of our model for Chhattisgarh}
    \label{tab:chhattisgarh-estimate}
\end{table}

Upon minimizing the error criterion \cref{eqn:LS-phi}, the estimated parameters in case of Chhattisgarh are obtained and presented in \cref{tab:chhattisgarh-estimate}. The fitted estimates for the time series variable of number of hospitalized $I_t$ and number of recoveries $R$ are shown in \cref{fig:chhattisgarh-fitted} to be compared against the actual observed values. Here, we use only the available data upto August 1, 2020 and use the estimated parameters to perform a short term prediction upto August 22, 2020 to check the validity of our estimation process. From \cref{fig:chhattisgarh-fitted}, it is clear that the estimation for the number of hospitalized and recovered patients seems sufficiently reasonable.

One of the most important quantity to consider in epidemiological studies is the reproduction number $R_0$. It denotes the number of secondary infection spread by an infected person on average. In our model, the population of infected individual is divided into three groups, $I_p, I_{sn}$ and $I_t$, all of whom has different transmission and recovery rates. In the model described in \cref{eqn:diff}, the reproduction number $R_0$ can be calculated as;

\begin{equation}
    R_0(t) = \dfrac{I_p(t)}{I_p(t) + I_{sn}(t) + I_t(t)} \left( \dfrac{\beta_{1t}(t)}{\theta_t (t) + \alpha + \lambda} \right) + \dfrac{I_{sn}(t)}{I_p(t) + I_{sn}(t) + I_t(t)} \left( \dfrac{\beta_{2t}(t)}{\delta_t (t) + \kappa + \zeta} \right)
    \label{eqn:R0}
\end{equation}

\begin{figure}[ht]
    \centering
    \includegraphics[width = \linewidth]{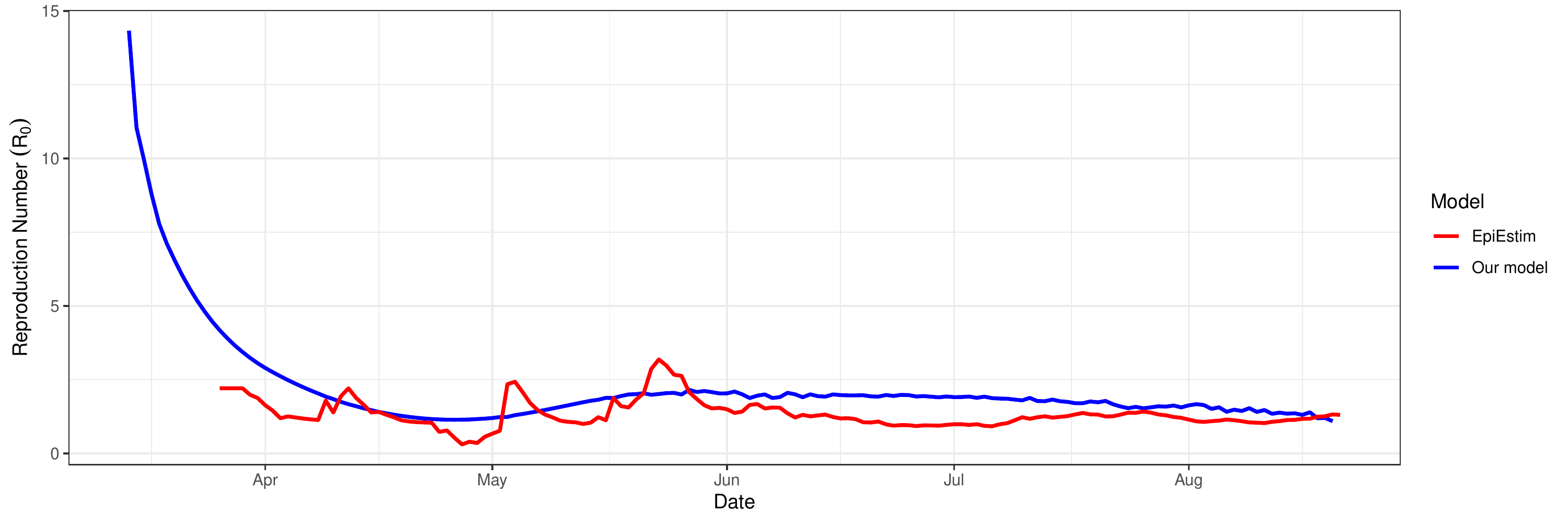}
    \caption{Time varying Reproduction number ($R_0$) estimated by our model and ``EpiEstim" package for Chhattisgarh}
    \label{fig:chhattisgarh-R0}
\end{figure}

The expression in~\cref{eqn:R0} basically takes the reproduction number corresponding to each of the group of infected individuals and then take a weighted average of them commensurate to the size of the groups. It should be noted that the transmission rate is assumed to be equal to $0$ (or negligible) in case of hospitalized patients on the basis of complete effectiveness of quarantine protocols. In order to validate our model estimation procedure, we use \texttt{EpiEstim} package~\cite{coriEpiEstim} in \texttt{R} to estimate time varying reproduction rate independently and compare the estimates against our estimates of $R_0$ for the state Chhattisgarh. Results are shown in~\cref{fig:chhattisgarh-R0}. Clearly, the prediction resembles closely overall, except in the month of July, where our estimate of $R_0$ is slightly higher than the estimates obtained using \texttt{EpiEstim} package. However, both estimates suggest an estimate of reproduction number about $1.25$ during the most recent period.

\begin{figure}[ht]
    \centering
    \includegraphics[width = \linewidth]{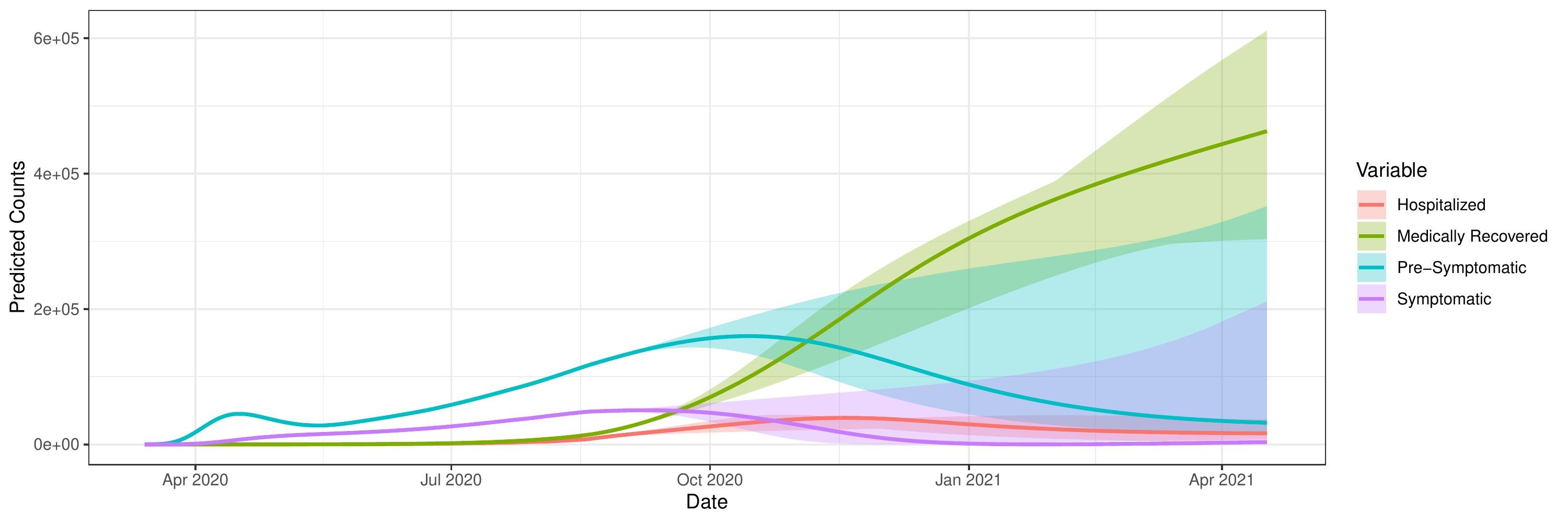}
    \caption{Long term prediction of pre-symptomatic, symptomatic, hospitalized, medically recovered (recorded) patients for Chhattisgarh along with $95\%$ confidence intervals}
    \label{fig:chhattisgarh-prediction}
\end{figure}

In order to be able to perform long term prediction, we use ARIMA model with seasonality to perform prediction of the time varying parameters like $\theta_t, \delta_t$. Akaike's Information criterion was chosen in order to obtain the best fitting ARIMA model. Also, in the current state of partial lockdown, it is reasonable to assume that the $\beta_{1t}$ is not likely to vary considerably in near future, and is, therefore, taken as the constant $b = 0.102$ throughout the period of prediction. Figure ~\ref{fig:chhattisgarh-prediction} shows the long term prediction for next $8$ months of the number of pre-symptomatic $(I_p)$, symptomatic $(I_sn)$, hospitalized $(I_t)$ and recorded recovered $(R)$ patients.

To compute a confidence interval for the predictions, we rely on the prediction confidence intervals obtained from the ARIMA model prediction of time varying parameters. For each of the time varying parameters, we consider the upper and lower bounds of the $95\%$ confidence interval for them. Then, for each combination of these boundary values, the model is simulated to the end of the prediction regime. With $4$ independent time varying parameters such as $\theta_t, \delta_t, c_2, c_3$,  and one dependent time varying parameter $c_1 = (1 - c_2 - c_3)$, we thus create $16$ such prediction scenarios, assuming other parameters to be fixed as their estimated values. Finally, the daywise minimum and maximum of all such prediction scenarios were taken in order to obtain an approximate $95\%$ confidence interval of the predictions.

Interestingly,~\cref{fig:chhattisgarh-prediction} shows that an indication of a small primary wave in April 2020, which then ends during middle of May, 2020. This could serve as an indication of the effectiveness of various lockdown enforcement schemes imposed by Govt. of India. However, the second wave starting from middle of May, 2020 could serve as an indication of increasing migrant movements in India, thus, creating the opportunity of more detrimental and imminent second wave in the virus spread. While it is clear that the number of pre-symptomatic infected individuals will rise considerably to approximately $2$ lakhs, the effect on the hospitalization rate will remain much lower, within a range of $25$-$30$ thousands.

We also perform a simple sensitivity analysis to find the importance of each of the parameters in prediction. The estimates of fatality Rate $\hat{\tau}$ and recovery rate $\hat{\gamma}$ turns out to be extremely robust and subtle changes to the estimates given in~\cref{tab:chhattisgarh-estimate} affects only the number of individuals in deceased state ($E$) and recovered state ($R$). The estimate of $\kappa$ affects primarily the $I_p$ and $I_{sn}$ states, and to some extent the prediction of $I_t$ as well. Increasing $\kappa$ to $0.15$ drops decreases $I_p$ and $I_{sn}$ both by $32\%$, while it slightly increases $I_t, R$ and $E$ states by $4\%$. The effect of $\alpha$ is fairly robust as long as $\alpha$ lies between $0.006$ and $0.018$ which incorporates about $50\%$ change over current estimate of $\alpha$. However, with $\alpha$ being relatively high like $0.05$ or more, the estimates of $I_{sn}$ and $I_t$ increase rapidly. We found that the estimates of $\lambda$ and $a$ are pretty sensitive, and these directly affect the shape of the incidence curve of $I_p$ and $I_{sn}$. We found that, about $7\%$ change of these estimates retains the similar shape with two phases of covid spread as shown in ~\cref{fig:chhattisgarh-prediction}. The estimate of $b$, and the migrant related parameters like $c_1, c_2, c_3$, has subtle effects on the change of $I_p$ and $I_t$. An increase in $b$ increases each of $I_p, I_{sn}$ and $I_t$ slightly and add a lagged effect towards increment of $U, E$ and $R$; An increase in $c_1, c_2$ increases $I_p$ and $c_3$ increases the incidence curve $I_t$.

The time varying parameters $\theta_t$ and $\delta_t$ (the testing rate for asymptomatic and symptomatic patients, respectively) have a similar significant contribution to the prediction. A change in $\delta_t$ is found to affect only the recorded hospitalized ($I_t$) and symptomatic ($I_{sn})$ cases, while a change in $\theta_t$ is found to correlate with hospitalized ($I_t$) and pre-symptomatic ($I_{p})$ cases. In case of Chattisgarh, increasing $\theta_t$ by $10\%$ reduces the peak of pre-symptomatic cases by about $7.2\%$ while increasing the peak of hospitalized cases by about $9.8\%$. However, as most of the detected covid cases are asymptomatic (pre-symptomatic) or mildly symptomatic in nature, an increase in $\delta_t$ turns out to be fairly robust, only decreasing $I_{sn}$ by $2.3\%$, and increasing $I_t$ even lesser, $1.3\%$. Furthermore, we have seen that a change in $\delta_t$ has minimal effect on the public health aspect of the disease, i.e. size of the peak, proportion of population affected etc. On the other hand, the size of the peak and the duration of the disease in the population is seen to be extremely sensitive to $\theta_t$. We elaborate on this more in the concluding section.

\subsection{Comparison with existing models} \label{comparison}

We have performed two comparative studies to see the performance in prediction of our model in comparison with the existing models in the literature like extended SIR model(eSIR) proposed by Song et al.~\cite{Song2020.02.29.20029421} and SIDARTHE model proposed by Giordano et al.~\cite{giordano2020modelling}.

\begin{figure}[ht]
    \centering
    \includegraphics[width = 0.49\linewidth]{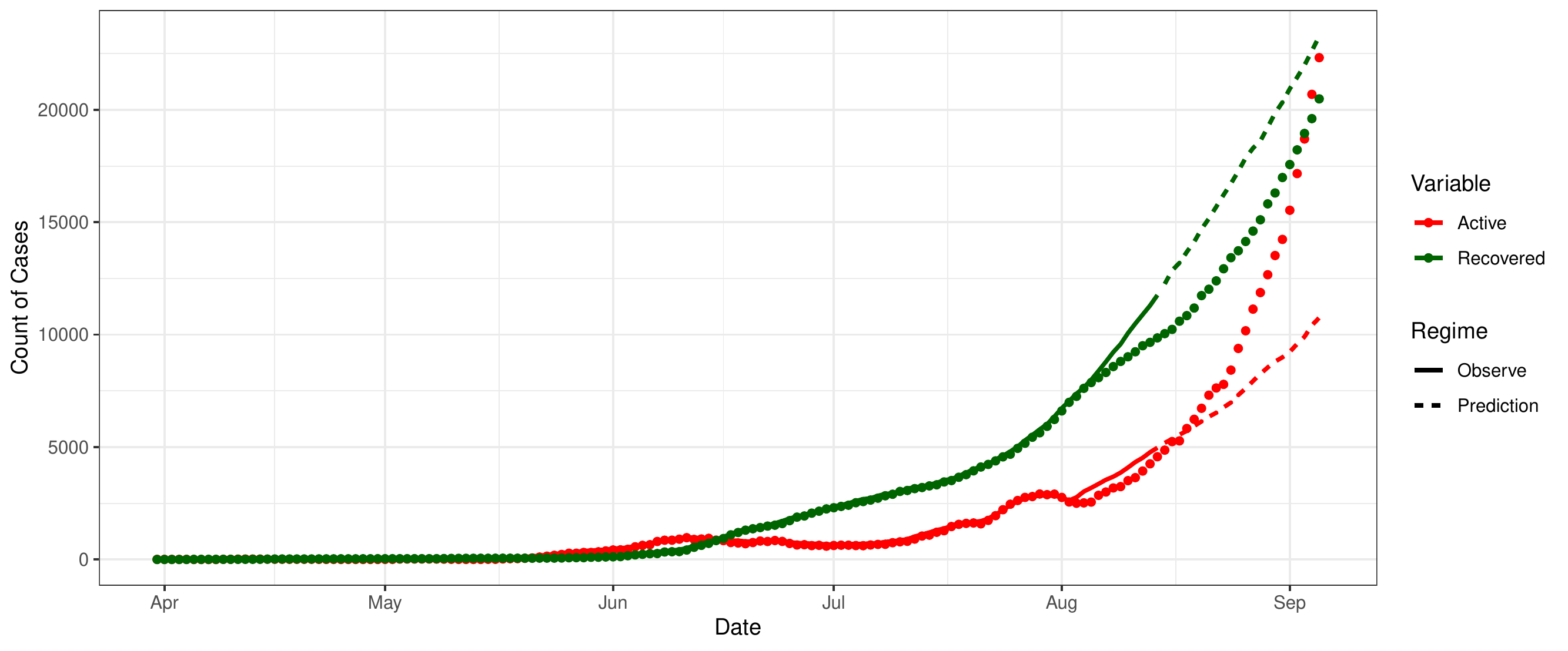}
    \includegraphics[width = 0.49\linewidth]{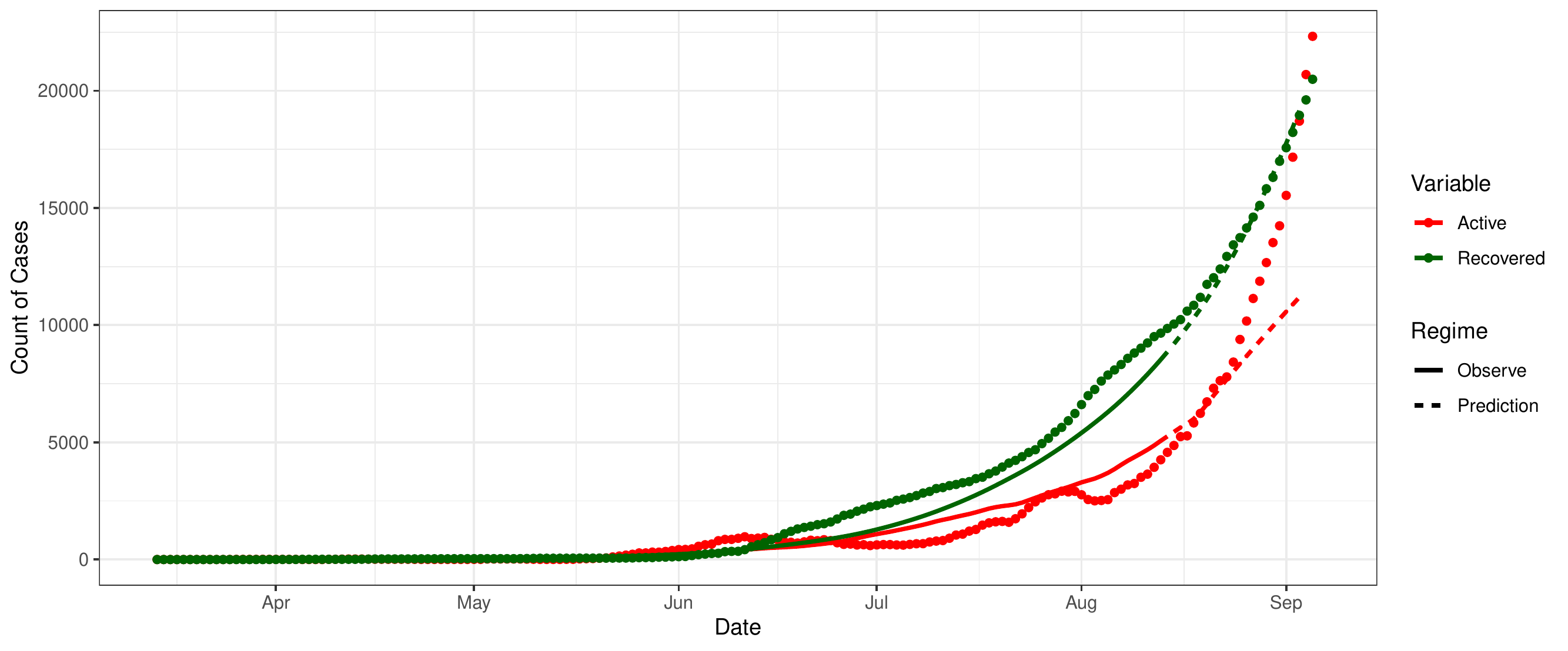}
    \caption{Prediction of trained eSIR(left) and SINTRUE(right) model from August 15, 2020 to September 5, 2020 for Chhattisgarh}
    \label{fig:chhattishgarh-eSIR-vs-sintrue}
\end{figure}

To compare the SINTRUE model with the eSIR model as presented in Song et al.~\cite{Song2020.02.29.20029421}, both models are fitted to the data from the state of Chhattisgarh upto August 15, 2020, and using the estimated parameters, the number of Reported Active cases and Recorded  Recovered cases upto September 5, 2020 were predicted. As a measure of deviation from the truth, Root Mean Square Error (RMSE) in prediction is used, and is found to be $4865.11$ for the prediction of Reported Active cases (corresponding to $I_t$ state in SINTRUE model), and $3174.49$ for the prediction of Recorded Recovered cases (corresponding to $R$ state in SINTRUE model) obtained by eSIR model. On the other hand, our SINTRUE model achieves much smaller RMSE $2946.312$ and $423.8527$ in prediction of Active and Recovered cases respectively. Figure~\ref{fig:chhattishgarh-eSIR-vs-sintrue} shows the comparative figures for predicted and observed cases for Active and Recovered cases for both the models. From Figure~\ref{fig:chhattishgarh-eSIR-vs-sintrue}, we see that eSIR model consistently underestimates the number of recorded active cases and overestimates the number of recorded recovered cases. In comparison, SINTRUE model depicts highly accurate prediction for recorded recovered cases, and slightly better prediction for recorded active cases, however, the problem of underestimation still persists.

\begin{figure}[ht]
    \centering
    \includegraphics[width = 0.49\linewidth]{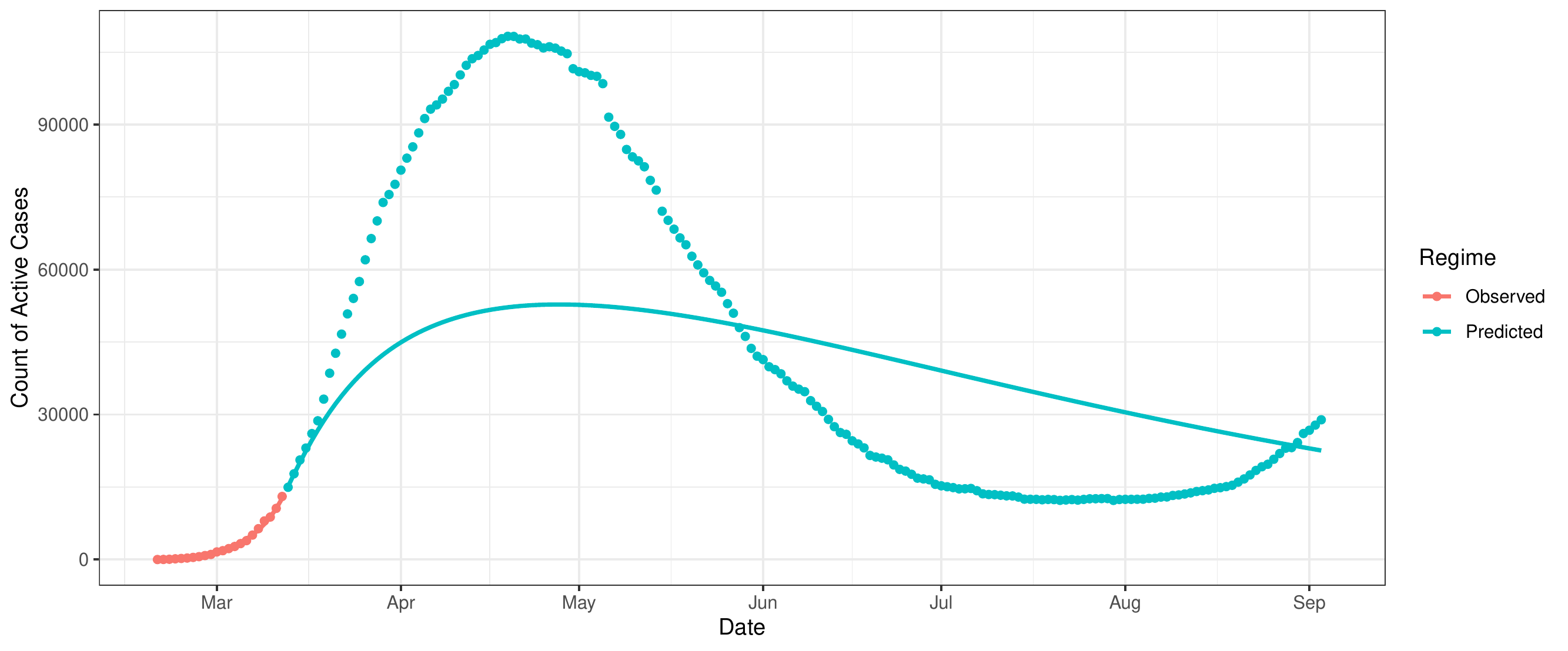}
    \includegraphics[width = 0.49\linewidth]{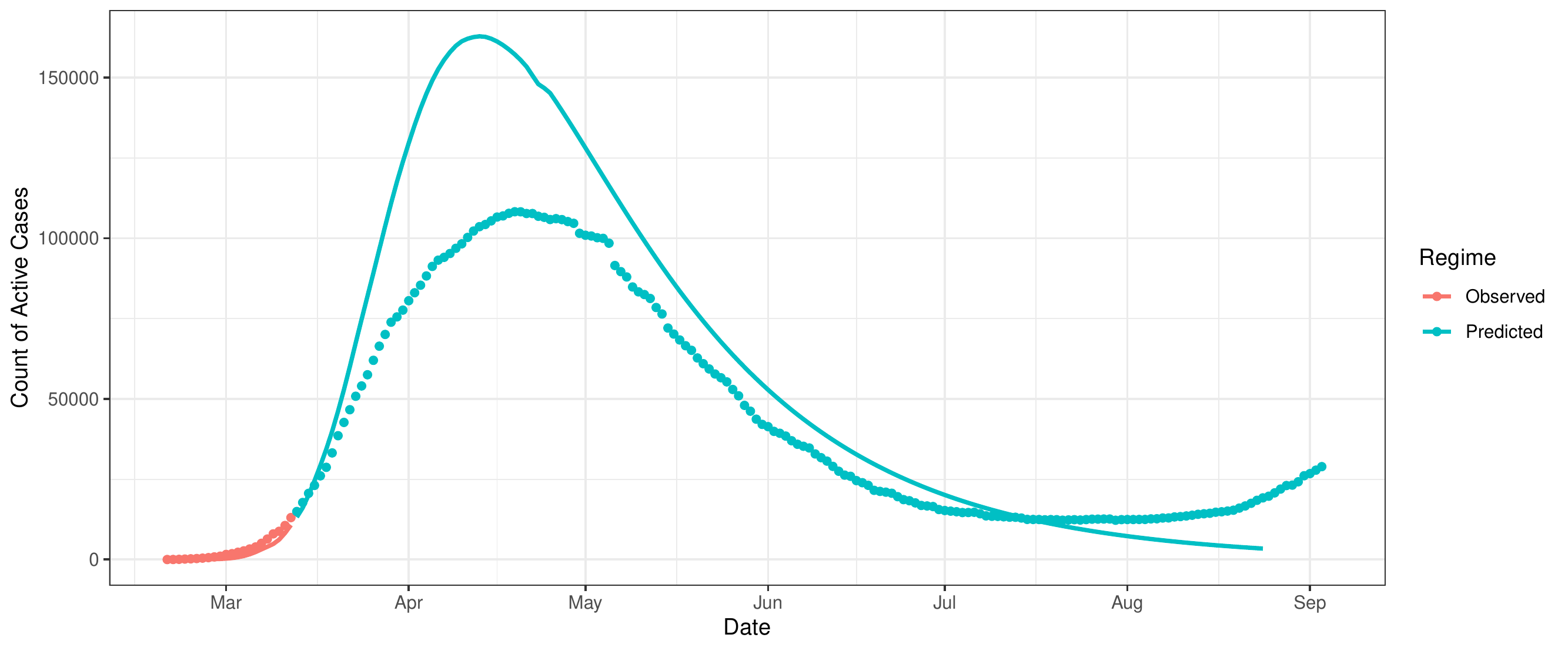}
    \caption{Prediction of SIDARTHE(left) and SINTRUE(right) model from March 13, 2020 to September 5, 2020 for Italy assuming no migration movement}
    \label{fig:italy-sidarthe-vs-sintrue}
\end{figure}

In order to assess comparative performance of the SIDARTHE model with SINTRUE model, we consider Italy's similar to Giordano et al.~\cite{giordano2020sidarthe-arxiv}. However, there was very little trustworthy data about migration situation in Italy~\cite{aless2020information}, which compelled us to reformulate SINTRUE model into a closed population model with $c_1 = c_2 = c_3 = 0$, and in particular $\diff{\text{In}} = \diff{\text{Out}} = 0$. With the data upto March 12 from various sources~\cite{dong19hongru, apicovid19italy} SINTRUE model with closed population is trained and a prediction is made from March 13, 2020 to September 5, 2020 for the active cases. In comparison, among the different scenarios for prediction presented in Giordano et al.~\cite{giordano2020sidarthe-arxiv}, we find that the scenario with slightly stronger lockdown and social distancing effect is found to yield a better alignment with the observed incidence curve, compared to the minimal social distancing scenario suggested in the paper. So, the parameter values presented in Giordano et al.~\cite{giordano2020sidarthe-arxiv} with mildly stronger social distancing is chosen. For the SIDARTHE model, the RMSE is found to be $27477.46$ while the RMSE with SINTRUE model's prediction is found to be slightly lesser $24768.48$. Figure ~\ref{fig:italy-sidarthe-vs-sintrue} reveal finer details. On the left panel, the SIDARTHE model starts by almost exact prediction, but soon diverges out, giving especially bad prediction near the tail. It also underestimates the peak of the pandemic, which might affect any policy decisions using this model. On the right panel, SINTRUE model provides an excellent estimate both at the start, i.e on short term basis and at the tail of the prediction curve, i.e on long term basis. Its nearly comparable MSE with the SIDARTHE model is accounted by the overestimation at the peak, which can be thought of as an upper bound to the peak number of affected cases, thereby helping policy formulations. Further, the SINTRUE model manages to altogether replicate the shape of the actual curve based on data only upto March 12, whereas the shape predicted by SIDARTHE model is not nearly correct. This is expected, since unlike the SINTRUE model, SIDARTHE model doesn't use any time-varying parameters, rendering it unsuitable for long-term predictions where the parameter values are liable to change.

\section{Conclusion}
The model presented in this paper comprises seven compartments in the progression of the disease, with the addition of an inflow to and an outflow of people from the population. Further, we have incorporated the pre-symptomatic~/asymptomatic population in the model as well as the population who get the virus but remain undetected throughout their journey from being infected to being recovered. The seven compartments considered in the model are \textbf{S}usceptible, \textbf{I}nfected and pre-symptomatic, Infected and Symptomatic but \textbf{N}ot Tested, \textbf{T}ested Positive, Recorded \textbf{R}ecovered, \textbf{U}nrecorded Recovered, and \textbf{E}xpired.

One extremely important observation that we make from the SINTRUE model and the subsequent simulation is, the testing rate of symptomatic patients actually does not affect the disease dynamics in any major way. Rather, it is the testing rate of the asymptomatic patients that turn out to be an extremely crucial parameter that can make or break the fight against the pandemic. The dynamic is extremely sensitive against the testing rate of the asymptomatic patients and once the rate goes up, the $R_0$ comes down drastically. The current $R_0$ indicates that around $23.664\%$ (of Chhattisgarh) of the population needs to be affected in order to  reach herd immunity~\cite{RANDOLPH2020737, kwok2020herd}. Hence, as a result of our model, one definitive suggestion we can make is that in order to fight the pandemic, one has to scale up their efforts on testing the asymptomatic patients. The increase in $\theta_t$ can be achieved by either increasing the $N_c$ in our model, which will mean an increase in the contact tracing endeavour, or to start allowing on-demand testing. Both of these strategies will increase $\theta_t$ and will bring down the peak considerably. As for the case of India, until September 4, India did not have a provision for people for getting themselves tested without any valid reason (symptoms, contacts etc.) and without being prescribed by a medical practitioner \cite{icmr-test-strategy-v5}. From September 4 onwards \cite{icmr-test-strategy-v6}, India has changed the testing strategy to permit anyone to get tested without any reason or prescription. We believe that this will have a direct positive impact on the asymptomatic testing rate $\theta_t$ and this can heavily affect the progression of the disease in a positive manner.

The advantage of the SINTRUE model is that the estimation for the parameter associated with extinct and recovery flow are pretty straightforward. Further, the estimation of the parameter to understand population movement from ``Asymptomatic" to ``Tested Positive" is obtained even without knowing the prevalence rate. This directly informs us that just looking at the rates of 'confirmed' cases, we cannot readily make a judgment about the prevalence of the disease in the population.

To understand the infection of susceptible populations from the asymptotic or symptomatic population, we choose the values $a$ and $b$.  The choice of parameters $\lambda$ (the rate of naturally healing population without getting detected), $a$, and $b$ control the shape of the incidence curve in a way that can give a realization of the second wave of COVID-19 or third wave scenarios. So, one advantage the model has is having a much broader range of incidence curves and from a realized incidence curve, a direct idea on the transmission rate can be arrived upon. On the downside, the grid search at the end requires heavy computation (although the codes may be optimized further). It certainly requires a good amount of granular data (like the recorded ration of pre-symptomatic and symptomatic patients, etc.). Based on the available data for one of India's states, i.e. Chhattisgarh, the results discussed in detail in $\S$~\ref{res} suggest that the parameters estimated with the migration and asymptomatic certainly can help in designing the COVID-19 control policy.  

The predictability of the model suggested in this paper is also compared (in $\S$~\ref{comparison}) with the the extended SIR model (eSIR) proposed by Song et al.~\cite{Song2020.02.29.20029421} and SIDARTHE model proposed by Giordano et al.~\cite{giordano2020modelling}. The predictability of proposed model in this paper is significantly better based on the performance measurement criteria of root mean squared error (RMSE) and the long-term predictability. The results obtained from the model represented in this paper have shown significantly better predictive capability compared to other newly developed models of COVID-19 progression, when applied on the data of Chhattisgarh State (India) and Italy. Further, this model could be thought out to extend as the aggregate model for the whole country, as the country when virtually under lock-down from May 2020, we can consider that on a state-wise aggregated form; the model can exhibit mass conservation property as $\diff{N}$ will essentially be $0$ for the country. We can show that given the initial states sum up to one, in the nationwide model's equilibrium state, the disease actually dies down. There will only be Susceptible, Recovered (documented or undocumented), and extinct people in the system at the equilibrium, which will mean that the epidemic (in the sense of the pandemic being treated only within the country) is over. This state will be reached eventually, but in order to speed that up, the recommendation from the SINTRUE model is that we have to change the testing policy to cover as much of the population as possible, and then only the spread can be arrested.

\section*{Acknowledgments}

\subsection*{Data}In this research, along with the data sources cited from open sources, the specific data to discuss the results of Chhattisgarh state, we acknowledge the support of Dr. Vinit Jain (Professor, Orthopaedic). Dr. Kamlesh Jain (Professor, Preventive Medicine) and Dr. Santosh Singh Patel (Associate Professor, Ophthalmology) from Pandit Jawahar Lal Nehru Memorial Medical College, Raipur.

\subsection*{Funding}
This work was supported by IIM Visakhapatnam under the seed research grant project titled ``Towards Prediction and Control of COVID- 19 in Particular and any Pandemic in General".


\end{document}